\documentclass[11]{article}

\usepackage{style/paper}

\title{Fault Tolerant and Fully Dynamic DFS in Undirected Graphs: Simple Yet Efficient}
\author{Surender Baswana
\thanks{
    Heinz Nixdorf Institute, University of Paderborn, Paderborn, Germany (https://www.hni.uni-paderborn.de),
    email: sbaswana@hni.upb.de}
\and Shiv Kumar Gupta
\thanks{
    Dept of Computer Science \& Engineering, Indian Institute of Technology Kanpur, Kanpur, India (https://www.cse.iitk.ac.in),
    email: \{shivg,atulsyan\}@cse.iitk.ac.in}
\and Ayush Tulsyan \footnotemark[2]
}
\date{}

\begin{document}

\maketitle

\begin{abstract}
We present an algorithm for a fault tolerant Depth First Search (DFS) Tree in an undirected graph. This algorithm is drastically simpler than the current state-of-the-art algorithms for this problem, uses optimal space and optimal preprocessing time, and still achieves better time complexity. This algorithm also leads to a better time complexity for maintaining a DFS tree in a fully dynamic environment.
%
%

\end{abstract}

\section{Introduction}
\label{sec:intro.tex}

Depth First Search (DFS) is a widely popular graph traversal method. The traversal routine, formalized by Tarjan~\cite{Tarjan72} in 1972, has played a crucial role in various graph problems including reachability, bi-connectivity, topological sorting, and strongly connected components.

Given an undirected graph~$G = \br{V, E}$ with~$n = |V|$ vertices and~$m = |E|$ edges, DFS traversal on the graph takes $\OO{m+n}$ time and results in a DFS tree (a rooted spanning tree) of~$G$. The traversal also classifies all the edges into tree edges and back edges.

Almost all real-world problems on graphs deal with updates. Vertices and edges keep entering and leaving the graph all the time. This update sensitive nature of graphs has motivated researchers to find methods for updating the solutions efficiently. There has been significant work in this area in the last two decades.

The {\em fault tolerant} version of any problem $\mathcal{P}$ on a graph $G$ is to construct a compact data structure, using which, for any set of failed edges or vertices $F$, one can efficiently report the solution of $\mathcal{P}$ on $G \setminus F$.
Many elegant fault tolerant algorithms have been designed for problems including connectivity~\cite{ChanPR08,Duan10,FrigioniI00}, shortest paths~\cite{BaswanaK13,ChechikLPR12,DemetrescuTCR08}, and spanners~\cite{BraunschvigCPS15,ChechikLPR10}.

The {\em dynamic} version of any problem $\mathcal{P}$ on $G$ is modeled as follows. For any online sequence of updates (insertion or deletion of edges/vertices), one has to report the solution of $\mathcal{P}$ efficiently after every update.
Note that, unlike the fault tolerant version, the updates are persistent in dynamic version, i.e., after each update, the solution has to be reported taking into account all the updates made so far. 
Algorithms which handle both insertion and deletion of vertices/edges are called {\em fully dynamic} graph algorithms, whereas the algorithms that handle either insertions or deletions are called {\em partially dynamic} graph algorithms, more specifically \textit{incremental} or \textit{decremental} graph algorithms, respectively.
The prominent results for dynamic graph problems include connectivity~\cite{EppsteinGIN97,HenzingerK99,HolmLT01,KapronKM13}, reachability~\cite{RodittyZ08,Sankowski04}, shortest path~\cite{DemetrescuI04,RodittyZ12}, matching~\cite{BaswanaGS18,BhattacharyaHI18,Solomon16}, spanner~\cite{BaswanaKS12,GottliebR08,Roditty12}, and min cut~\cite{Thorup07}.


\subsection{Previous Results on Fault Tolerant and Dynamic DFS}
Apart from the hardness results for the dynamic \textit{oredred} DFS problem (Reif ~\cite{Reif85,Reif87}, Miltersen \etal.~\cite{MiltersenSVT94}), little contribution was made until recently.
For directed acyclic graphs (DAG), Franciosa~\etal.~\cite{FranciosaGN97} presented an algorithm for maintaining a DFS tree under incremental edge updates which takes overall $\OO{mn}$ time for any sequence of $m$ edge insertions.
For undirected graphs, Baswana and Khan~\cite{BaswanaK14} in 2014 presented an incremental algorithm for a DFS tree which takes overall $\OO{n^2}$ time for any sequence of edge insertions.
Baswana and Choudhary~\cite{BaswanaC15} designed a randomized decremental algorithm for a DFS tree in DAGs with overall expected $\OO{mn\log{n}}$ time over any sequence of edge deletions.

None of the partially dynamic algorithms stated above achieves an $o\br{m}$ bound over the worst-case complexity of a single update.
Moreover, there were no fully dynamic or fault tolerant algorithms for DFS in undirected graphs until recently. In 2016, the first fault
tolerant algorithm was presented that takes $O(nk \log^4 n)$ time to report a DFS tree for any set of $k$ failed vertices or edges \cite{BaswanaCC016}. The time complexity was further improved by Chen \etal.~\cite{ChenDWZ16} to $\OO{nk \log^2 n}$. Both \cite{BaswanaCC016, ChenDWZ16} require a data structure occupying $\OO{m\log^2 {n}}$ bits.
Nakamura and Sadakane~\cite{NakamuraS17} reduced the space occupied by the data structure to $\OO{m\log{n}}$ bits which is indeed optimal \footnote{Precisely, their data structure occupies $\br{m+o\br{m}}\log{n}$ bits by using a wavelet tree.}. The fault tolerant algorithms presented in \cite{BaswanaCC016, ChenDWZ16, NakamuraS17} lead to $o\br{m}$ fully dynamic algorithms for maintaining a DFS tree using the standard technique of periodic rebuilding (refer to Table \ref{tab:comparison}). Recently, Chen \etal.~\cite{ChenDWZZ18} designed an $\OO{n}$ time incremental algorithm for DFS tree in undirected graphs, which is optimal if it is required to output the DFS tree after each update.

\subsection{Familiarizing with the Problem}
\label{subsec:problem-statement}
\label{prop:DFS}
For an undirected graph, the depth first nature of the traversal ensures that there are no cross edges in the resulting DFS tree. Following lemma states this property (referred to as \textit{DFS Property} henceforth):
\begin{lemma}
    For each vertex $v\in V$, a neighbour of $v$ appears either as an ancestor of $v$ or as a descendant of $v$ in any DFS tree.
    \label{lem:dfs}
\end{lemma}
We now define the ancestor-descendant path.
\begin{definition}(\textit{Ancestor-descendant path})
    A path in a DFS tree is called an ancestor-descendant path if its endpoints have an ancestor-descendant relationship in the tree.
    \label{def:anc-des-path}
\end{definition}

In the presence of failures of vertices or edges, the aim is to update the tree efficiently such that the DFS property is restored. In order to familiarize the reader with the problem, let us consider another related, but simpler problem, namely, rerooting of a DFS tree defined as follows.

\begin{problem}
    \label{problem:reroot}
    Preprocess an undirected graph $G=(V,E)$ to build a compact data structure so that given any vertex $v\in V$, we can report the DFS tree rooted at $v$ efficiently.
\end{problem}
\tikzstyle{vertex}=[
    circle,
    draw,
    solid,
    fill=black,
    scale=0.6,
    minimum size=1em
]
\tikzstyle{triangle}=[
    draw=black,
    fill=violet!24,
    regular polygon,
    regular polygon sides=3,
    node distance=1cm,
    minimum height=4em
]
\tikzstyle{emptyTriangle}=[
    fill=white,
    regular polygon,
    regular polygon sides=3,
    node distance=1cm,
    minimum height=4em
]
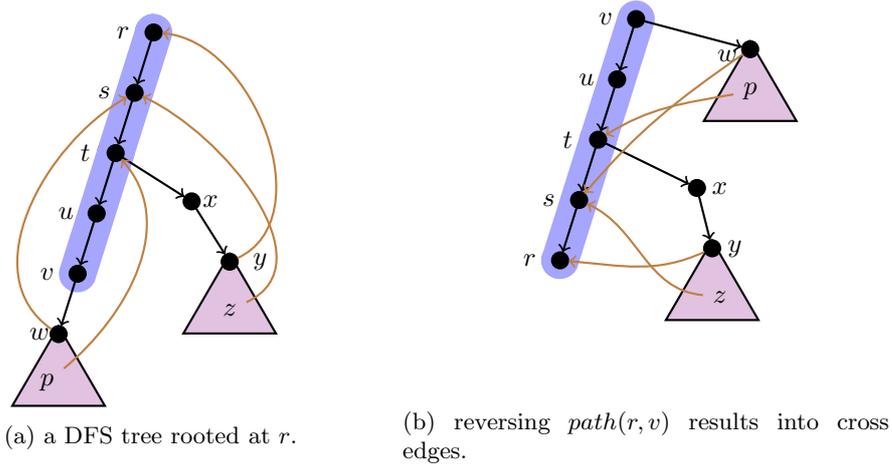
\begin{figure*}
    \centering
    \begin{subfigure}{.40\textwidth}
        \centering
        \begin{tikzpicture}[thick]
            \node (r) at (0, 0*0.8) [vertex] {};
            \node (r_l) at (-0.4, 0*0.8) {$r$};
            \node (s) at (0-.25, -1*0.8) [vertex] {};
            \node (s_l) at (-0.65, -1*0.8) {$s$};
            \node (t) at (0-.5, -2*0.8) [vertex] {};
            \node (t_l) at (-0.90, -2*0.8) {$t$};
            \node (u) at (0-.75, -3*0.8) [vertex] {};
            \node (u_l) at (-1.15, -3*0.8) {$u$};
            \node (v) at (-1, -4*0.8) [vertex] {};
            \node (v_l) at (-1.4, -4*0.8) {$v$};
            \node (tw) at (0-1.25,-5.75*0.8) [triangle] {};
            \node (w) at (0-1.25, -5*0.8) [vertex] {};
            \node (w_l) at (-1.5, -5*0.8) {$w$};
            \node (p_l) at (-1.4, -5.8*0.8) {$p$};

            \node (x) at (.5, -2.8*0.8) [vertex] {};
            \node (x_l) at (.75, -2.8*0.8) {$x$};
            \node (ty) at (1,-4.55*0.8) [triangle] {};
            \node (y) at (1, -3.8*0.8) [vertex] {};
            \node (y_l) at (1.4, -3.8*0.8) {$y$};
            \node (z_l) at (1, -4.6*0.8) {$z$};

            \draw[->]  (r) -- (s);
            \draw[->]  (s) -- (t);
            \draw[->]  (t) -- (u);
            \draw[->]  (u) -- (v);
            \draw[->]  (v) -- (w);
            \draw[->]  (t) -- (x);
            \draw[->]  (x) -- (y);

            \begin{pgfonlayer}{background}
                \fill[blue,opacity=0.35] \convexpath{r,t,v,t,r}{7pt};
            \end{pgfonlayer}

            \draw[->] [brown,line width=0.8pt] (p_l) to[out=40,in=312] (t);
            \draw[->] [brown,line width=0.8pt] (z_l) to[out=25,in=335] (s);
            \draw[->] [brown,line width=0.8pt] (y) to[out=25,in=355] (r);
            \draw[->] [brown,line width=0.8pt] (w) to[out=145,in=210] (s);

        \end{tikzpicture}
        \caption{ a DFS tree rooted at $r$.}
    \end{subfigure}
    \begin{subfigure}{.40\textwidth}
        \centering
        \begin{tikzpicture}[thick]
            \node (v) at (0, 0*0.8) [vertex] {};
            \node (v_l) at (-0.4, 0*0.8) {$v$};
            \node (u) at (0-.25, -1*0.8) [vertex] {};
            \node (u_l) at (-0.65, -1*0.8) {$u$};
            \node (t) at (0-.5, -2*0.8) [vertex] {};
            \node (t_l) at (-0.9, -2*0.8) {$t$};
            \node (s) at (0-.75, -3*0.8) [vertex] {};
            \node (s_l) at (-1.15, -3*0.8) {$s$};
            \node (r) at (-1, -4*0.8) [vertex] {};
            \node (r_l) at (-1.4, -4*0.8) {$r$};
            \node (tw) at (1.5,-1.25*0.8) [triangle] {};
            \node (w) at (1.5, -.5*0.8) [vertex] {};
            \node (w_l) at (1.2, -.6*0.8) {$w$};
            \node (p_l) at (1.5, -1.2*0.8) {$p$};
            \node (dummy) at (0-1.25,-5.75*0.8) [emptyTriangle] {};

            \node (x) at (.8, -2.8*0.8) [vertex] {};
            \node (x_l) at (1.1, -2.8*0.8) {$x$};
            \node (ty) at (1,-4.55*0.8) [triangle] {};
            \node (y) at (1, -3.8*0.8) [vertex] {};
            \node (y_l) at (1.3, -3.8*0.8) {$y$};
            \node (z_l) at (1.1, -4.6*0.8) {$z$};

            \draw[->]  (v) -- (u);
            \draw[->]  (u) -- (t);
            \draw[->]  (t) -- (s);
            \draw[->]  (s) -- (r);
            \draw[->]  (v) -- (w);
            \draw[->]  (t) -- (x);
            \draw[->]  (x) -- (y);

            \begin{pgfonlayer}{background}
                \fill[blue,opacity=0.35] \convexpath{r,t,v,t,r}{7pt};
            \end{pgfonlayer}

            \draw[->] [brown,line width=0.8pt] (p_l) to[out=190,in=32] (t);
            \draw[->] [brown,line width=0.8pt] (z_l) to[out=175,in=335] (s);
            \draw[->] [brown,line width=0.8pt] (y) to[out=205,in=355] (r);
            \draw[->] [brown,line width=0.8pt] (w) to[out=215,in=50] (s);

        \end{tikzpicture}
        \caption{reversing $path(r,v)$ results into cross edges.}
        \label{fig:cross}
    \end{subfigure}
    \caption{Non-triviality of rerooting problem.}
    \label{fig:reroot}
\end{figure*}

Let $T$ be an initial DFS tree, rooted at a vertex, say $r$. For computing a DFS tree rooted at any vertex $v$, the most obvious solution may be
to just reverse the direction of the path from $r$ to $v$ in $T$. However, this may result in transforming many back edges to cross edges and hence a violation of the DFS property (see Figure \ref{fig:reroot}). To fix this problem, we may need to reroot various subtrees hanging from the reversed path. Along these lines,~\cite{BaswanaCC016} presented an algorithm that takes $O(n\log ^3 n)$ time to compute a DFS tree rooted at any vertex.

In order to see how rerooting a DFS tree is related to the problem of fault tolerant DFS tree, consider the failure of vertex $u$ in Figure \ref{fig:reroot}(a). The subtree $T(v)$ is connected to the remaining tree through many back edges and $(p,t)$ is the lowest among all of them. If we reroot the subtree $T(v)$ at vertex $p$ and hang it from the remaining tree through the edge $(p,t)$, this provides a valid DFS tree of $G\backslash \{u\}$.

The problem of fault tolerant DFS tree becomes more complex in the presence of multiple faults. However, the distribution of faults plays an important role as follows.
If the failing vertices do not have any ancestor-descendant relationship in the DFS tree, they can be handled {\em independently}. For example, the simultaneous failure of
vertices $u$ and $x$ in the DFS tree shown in Figure \ref{fig:reroot}(a) requires rerooting the respective subtrees $T(v)$ and $T(y)$ at vertices $p$ and $z$ respectively.
So, even for arbitrarily large number of failures, if none of them have the ancestor-descendant relationship, we have to just reroot the corresponding disjoint subtrees of $T$ to report the DFS tree avoiding those failures.
If two or more failures indeed have the ancestor-descendant relationship,~\cite{BaswanaCC016} presents an algorithm which is quite different from their rerooting algorithm.

\subsection{Overview of the Previous Results}
We now begin with an overview of the existing algorithms for fault tolerant DFS tree. For any set of $k$ failures, the algorithm presented in~\cite{BaswanaCC016} first partitions the original DFS tree into a pool of connected components. This pool consists of $k$ paths (specifically, \textit{ancestor-descendant paths}) and potentially a large number of subtrees. The algorithm treats each of these components as a {\em super vertex} and uses them to grow the DFS tree $T^*$ that avoids all the failures.
At a high level, the algorithm can be visualized as a traversal on these super vertices. Each traversal extracts a path from the super vertex, attaches it to $T^*$, and places the remaining portion of the super vertex back into the pool. In order to pursue DFS traversal further in an efficient manner, the algorithm needs to compute \textit{minimal} adjacency lists for the vertices of the traversed path (referred to as \textit{reduced adjacency lists}).
The algorithm makes use of the following crucial property of DFS traversal.
\begin{lemma}
    \label{lem:components}
    [Components Property~\cite{BaswanaCC016}] Consider any DFS Traversal on any undirected graph $G = \br{V, E}$. When the traversal reaches a vertex $v \in V$, let the set of connected components induced by the unvisited vertices be $C$. If from any component $c \in C$, there exists two edges - $e$ to vertex $v$ and $e'$ to any of the visited vertices (including $v$), then for building a valid DFS tree, it is sufficient to consider only the edge $e$ during the rest of DFS traversal, that is, $e'$ can be ignored.
\end{lemma}

In order to use the above property to populate the reduced adjacency list,
the algorithm needs a data structure to answer the following queries repeatedly.
\begin{itemize}
    \item $Query(w,x,y)$: among all the edges from $w$ that are incident on the $path(x,y)$ in $T$, return an edge that is incident nearest to $x$ on the $path(x,y)$.
    \item $Query(T(w),x,y)$: among all the edges from $T(w)$ that are incident on the $path(x,y)$ in $T$, return an edge that is incident nearest to $x$ on the $path(x,y)$.
\end{itemize}
It is quite obvious from the description given above that
these queries are quite non-trivial, and so a sophisticated data structure is designed in \cite{BaswanaCC016} to answer these queries efficiently. 
In addition to the complex data structure, the complete difference in the processing of a path and a subtree obfuscates the algorithm and its analysis.

The subsequent results \cite{ChenDWZ16, NakamuraS17} keep the algorithm unchanged and replace the data structure used in~\cite{BaswanaCC016} with alternate data structures.
Chen \etal.~\cite{ChenDWZ16} model the two queries mentioned above as Orthogonal Range Successor/Predecessor(ORS/ORP) queries and this improves the query processing time.
Nakamura and Sadakane~\cite{NakamuraS17} compressed the data structure used in~\cite{ChenDWZ16} using Wavelet Trees~\cite{GrossiGV03} to achieve optimal space.
Despite these improvements, the core of the fault tolerant algorithm remains intricate and the data structure still remains complex. Recently, in an empirical study \cite{BaswanaG018}, it was found that this algorithm performs even worse than the static DFS algorithm for certain classes of graphs.
This naturally raises the question if there exists a simpler algorithm for this fundamental problem.



Moreover, all these algorithms fail to incorporate the distribution of the faults to achieve efficiency - Consider the case where very few among the $k$ faults in the tree have an ancestor-descendant relationship, these algorithms process this input in a way similar to the input with worst case distribution of the faults and take $O(nk \text{ } polylog(n))$ time although there is a simple $O(n \text{ } polylog(n))$ time algorithm for such cases as described in \ref{subsec:problem-statement}.

\subsection{Our Contribution}

We take a much simpler approach as compared to the previous algorithms. 
We decompose the DFS tree $T$ into a disjoint collection ${\cal P}$ of ancestor-descendant paths. Similar to~\cite{BaswanaCC016}, each of these paths are treated like {\em super vertices}. At a high level, the algorithm can still be viewed as a traversal on these super vertices.
However, as the reader may also verify, the algorithm turns out to be lighter and quite different at the core. Interestingly, the original DFS tree alone acts as a powerful data structure to be used for rerooting or for the computation of another valid DFS tree in the presence of faults.
The algorithm crucially exploits an implicit hierarchy among the ancestor-descendant paths in ${\cal P}$. This hierarchy along with the DFS property of $T$ enables us to use much simpler queries. In particular, each query will ask only for an edge from a vertex to one of its ancestor paths in the hierarchy. The hierarchy allows us to represent $T$ as another tree like structure, called {\em shallow tree}. In a nutshell, our algorithm can be viewed as an efficient DFS traversal guided by this shallow tree.

We first present a new and simple rerooting algorithm based on the above ideas.
This rerooting algorithm extends to the fault tolerant algorithm with very little and obvious modifications. While preserving simplicity, the fault tolerant algorithm turns out to be faster than all the previous algorithms. Moreover, our algorithm is the first to implicitly incorporate the distribution of faults to gain efficiency. We summarize our result in the following theorem.

\begin{theorem}
    An undirected graph $G$ and its DFS tree $T$ can be preprocessed in $\OO{m+n}$ time to build a data structure of $\OO{m+n}$ size, using which one can compute the DFS tree of the graph for any given $k$ failed vertices or edges, in $\OO{n\br{k'+\log{n}}\log{n}}$ time, where $k' \leq k$ is the maximum number of faults on any root-leaf path in the tree $T$.
    \label{thm:main}
\end{theorem}

\noindent
We now present the highlights of our algorithm.

\noindent
{\em Drastically simpler algorithm:}~Our algorithm is drastically simpler and more intuitive than the previous algorithm. We feel confident to defend that it can be taught even in an undergraduate course on algorithms. The pseudo-codes in Algorithm~\ref{algo:reroot} and Algorithm~\ref{algo:reducedal} are concise and very close to the corresponding implementations.
\\

\noindent
{\em Faster time complexity:}~Our algorithm takes $O(n(k'\log n + \log^2 n))$ time, where $k'$ is the maximum number of failures on any ancestor-descendant path of the DFS tree when $k$ edges/vertices fail. In the worst-case $k'$ can be as large as $k$. However, $k'$ can be $o(k)$ as well.
In the latter case, our result improves all the existing results significantly.
Moreover, even in the case $k'=k$, our time complexity is superior to the previous best by a $\log$ factor. \\

\noindent
{\em Optimal preprocessing time:}~ Our preprocessing relies upon DFS traversal only, taking $\OO{m+n}$ time. Given a graph, in order to report the initial DFS tree, one anyway has to run a static DFS. Hence, our preprocessing time is optimal.
\\

\noindent
{\em Optimal space and elementary data structure:}~In contrast to the heavy data structure used by~\cite{BaswanaCC016,ChenDWZ16}, our algorithm makes use of very elementary data structures which are compact as well. Each vertex keeps array storing edges incident on it from ancestors sorted according to their levels. This data structure uses just $m$ words and still achieves $O(n (k'\log^2 n + \log^3 n))$ time to report a DFS tree upon failure of any $k$ vertices or edges. By using fractional cascading~\cite{ChazelleG86}, we get rid of one $\log$ factor while still keeping space requirement to be $O(m)$.\\

\noindent
{\em Faster Fully Dynamic Algorithm:}~ Using Theorem \ref{thm:main} and periodic rebuilding technique used in \cite{BaswanaCC016},
we also get the fastest (till date) algorithm for fully dynamic DFS: \par
\begin{theorem}
    Given an undirected graph, one can maintain a DFS tree for any online sequence of insertions and deletions of vertices/edges in $\OO{\sqrt{mn\log{n}}}$ worst-case time per update.
\end{theorem}
The new fully dynamic algorithm can be used to solve the dynamic subgraph problems discussed in~\cite{BaswanaCC016} and improves upon their time complexity as well (see Appendix).

Table 1 offers a comparison of our results with all the previous results.

\begin{table}[h!]
    \center
    \renewcommand{\arraystretch}{1.75}
    \caption{Comparison of the existing results and the new algorithm}
    \label{tab:comparison}
    \begin{tabular}{|l|c|c|c|c|}
        \hline
        & \cite{BaswanaCC016} & \cite{ChenDWZ16} & \cite{NakamuraS17} & New\\ \hline
        Space bits & $\mathcal{O}(m\log^2{n})$ & $\mathcal{O}(m\log^2{n})$ & $\mathcal{O}(m\log{n})$ & $\mathcal{O}(m\log{n})$\\ \hline
        Preprocessing & $\mathcal{O}(m\log{n})$ & $\mathcal{O}(m\log{n})$ & $\OO{m\sqrt{\log{n}}}$ & $\OO{m+n}$\\ \hline
        $k$-fault tolerant & $\mathcal{O}(nk\log^4{n})$ & $\mathcal{O}(nk\log^2{n})$ & $\mathcal{O}(nk\frac{\log^3{n}}{\log{\log{n}}})$ & $\OO{n(k' + \log{n})\log{n}}$\\ \hline
        Dynamic DFS & $\mathcal{O}(\sqrt{mn}\log^{2.5}n)$ & $\mathcal{O}(\sqrt{mn}\log^{1.5}{n})$ & $\mathcal{O}(\sqrt{mn}\frac{\log^{1.75}{n}}{\sqrt{\log{\log{n}}}})$ & $\mathcal{O}(\sqrt{mn\log{n}})$ \\ \hline
    \end{tabular}
\end{table}

\subsection{Organisation of the Paper}
We now briefly present how this paper is organized. Section \ref{sec:prelim} introduces the notations and some well-known techniques/properties used throughout the paper.
Section \ref{sec:shallow-tree} first defines the {\em shallow tree} representation, a concise structure which encapsulates the hierarchy of paths in the initial DFS tree. Section \ref{sec:reroot} is the core of our work. Here, we describe how a DFS tree can be rerooted efficiently.
Section \ref{sec:k-fault} describes how with some minor modifications to the data structure, rerooting procedure extends to a fault tolerant algorithm. We present the fully dynamic algorithm and its application to various dynamic subgraph problems in Section \ref{sec:fully-dynamic}.

\section{Preliminaries}
\label{sec:prelim}

\subsection{Notations}

Following notations will be used throughout this paper.
\begin{itemize}
    \item $T$: Any DFS tree of the original graph $G$.
    \item $path\br{x, y}$: the path from vertex $x$ to $y$ in $T$.
    \item $dfn\br{x}$:~The depth first number, i.e., the number at which vertex $x$ is visited during the DFS traversal.
    \item $v\br{i}$: the vertex $x\in V$ such that $dfn(x)=i$.
    \item $T\br{x}$: the subtree of $T$ rooted at vertex $x$.
\end{itemize}

For the sake of ease of explanation, we shall assume that the graph remains connected at all times. This assumption is
without loss of generality because of the following standard way of transforming the original graph right in the beginning -
Introduce a dummy vertex $r$ and connect it to all vertices of the graph. Henceforth, we maintain a DFS tree rooted at $r$ for this augmented graph. It is easy to observe that the augmented graph remains connected throughout and the DFS tree rooted at $r$
will be such that the subtrees rooted at the children of $r$ constitute the DFS forest of the original graph.




\subsection{Heavy-Light Decomposition}
Sleator and Tarjan, in their seminal result on dynamic trees~\cite{SleatorT83} introduced a technique of partitioning any rooted tree called Heavy-light decomposition.
Given any rooted tree, this technique splits it into a set of vertex-disjoint ancestor-descendant paths. It marks all the tree edges either dashed or solid - a tree edge is marked solid iff the subtree of the child vertex is heaviest among the subtrees of all its siblings and dashed otherwise. A maximal sequence of vertices connected through solid edges constitutes the required ancestor-descendant path. This decomposition can be carried out easily in $\OO{n}$ time.


\subsection{Fractional Cascading}
Given $n$ sorted arrays and a value $x$, suppose we need to find the predecessor/successor of $x$ in each of them. A naive way is to make a binary search on each array. Chazelle and Guibas~\cite{ChazelleG86} introduced a novel tool called fractional cascading using which this problem can be solved more efficiently.
Also, Chen \etal.\cite{ChenDWZZ18} used this tool for arriving at an $\OO{n}$ algorithm for incremental updates. We adapt a customized version of their method.

\begin{lemma}
    \label{lem:FC}
     Fractional cascading: Given $n$ sorted arrays $\{A_i\}_{i \in [n]}$ each with size $l_i = |A_i|$ and cumulative size, $ \sum_i^n l_i = m$, there exists an $\OO{m}$ space data structure which can be built in $\OO{m}$ time, such that for any given $x, i,$ and $k$ satisfying $i, k \in [n]$ and $i + k \leq n$, we can search for $x$ (or its predecessor/successor) in all arrays $A_i, \dots, A_{i+k}$ using the data structure in $O(k + \log{m})$ time.
\end{lemma}
\noindent
For the sake of completeness, we provide a description about the structure in Lemma \ref{lem:FC}.

Build $n$ sorted arrays $\{\mathcal{F}_i\}_{i \in [n]}$ as follows.
$\mathcal{F}_1$ is a copy of $A_1$. For $i>1$, $\mathcal{F}_i$ will be built by merging $A_i$ and $\mathcal{F}_{i-1}^{even}$. Here, $\mathcal{F}_{t}^{even}$ represents the array consisting of the elements at even indices in $\mathcal{F}_{t}$, i.e., $\mathcal{F}_{t}^{even}[j] = \mathcal{F}_{t}[2j] \hspace{2ex} \forall j \in [\lfloor n/2 \rfloor ]$. Note that all $\mathcal{F}_{i}$'s remain sorted after the merge. The number of elements in this structure is at most $2m$.

For each element $v$ in any array $\mathcal{F}_i$, we store two pointers, one to $A_i$ and one to $\mathcal{F}_{i-1}^{even}$. They point to the elements in the corresponding array that is just larger than or equal to $v$.

To answer the query for any triplet of integers $x$, $i$, and $k$, first search for $x$ in $\mathcal{F}_{i+k}$. Pointer to $A$ can be used to find the required element in $A_{i+k}$. Now instead of making another binary search in $\mathcal{F}_{i+k-1}$, pointer to $\mathcal{F}_{i+k-1}^{even}$ stored at the element in $\mathcal{F}_{i+k}$ is utilized. This leads us to the required element (or its immediate neighbours) in $\mathcal{F}_{i+k-1}$. By comparing $x$ with the element at pointed location (and its immediate neighbours), one can reach where another binary search would have led us. Here, again pointers to $A_{i+k-1}$ and $\mathcal{F}_{i+k-2}^{even}$ are available. This process can be repeated till one reaches $\mathcal{F}_i$.

\section{Shallow Tree Representation}
\label{sec:shallow-tree}



We now introduce the shallow tree representation for DFS tree $T$ that plays a key role in our algorithm. Using heavy-light decomposition, $T$ is broken down into a set of vertex-disjoint ancestor-descendant paths. Let's denote this set with $\mathcal{P}$. Observe that these paths are connected through dashed edges in $T$. These edges introduce a hierarchy among paths in $\mathcal{P}$ and
the shallow tree defined below captures this hierarchy.

\begin{definition}
    \label{def:shallow-tree}
    Given a DFS tree $T$ of an undirected graph $G$, let $\mathcal{P}$ be the set of paths obtained through heavy-light decomposition of $T$. Let $H$ be the set of edges marked dashed during the decomposition. For tree $T$, its shallow tree $S$ is a rooted tree formed by collapsing each element of ${\cal P}$ into a single node (super vertex). Note that, for each edge $(y,z)\in H$ with $y=parent(z)$, the node in $S$ that contains $y$ is the parent of the node containing $z$.
\end{definition}
\tikzstyle{vertex}=[
    circle,
    draw,
    solid,
    fill=white,
    inner sep=2pt,
    minimum size=1.5em
]
\tikzstyle{emptyVertex}=[
    circle,
    solid,
    fill=white,
    inner sep=2pt,
    minimum size=1.5em
]
\begin{figure*}
    \centering
    \begin{subfigure}{.30\textwidth}
        \centering
        \begin{tikzpicture}[thick]
            \node (a) at (0*0.8, 0*0.8) [vertex] {$a$};
            \node (b) at (-0.5*0.8,-2*0.8) [vertex] {$b$};
            \node (c) at (-1*0.8,-4*0.8) [vertex] {$c$};
            \node (d) at (-1.5*0.8,-6*0.8) [vertex] {$d$};
            \node (e) at (-0.5*0.8,-6*0.8) [vertex] {$e$};
            \node (f) at (0.5*0.8,-4*0.8) [vertex] {$f$};
            \node (g) at (0.5*0.8,-6*0.8) [vertex] {$g$};
            \node (h) at (2*0.8,-2*0.8) [vertex] {$h$};
            \node (i) at (2*0.8,-4*0.8) [vertex] {$i$};
            \node (j) at (1.5*0.8,-6*0.8) [vertex] {$j$};
            \node (k) at (2.5*0.8,-6*0.8) [vertex] {$k$};
            \node (l) at (3*0.8,-2*0.8) [vertex] {$l$};

            \draw [->] (a) -- (b);
            \draw [->] (b) -- (c);
            \draw [->] (c) -- (d);
            \draw [->] (c) -- (e);
            \draw [->] (b) -- (f);
            \draw [->] (f) -- (g);
            \draw [->] (a) -- (h);
            \draw [->] (h) -- (i);
            \draw [->] (i) -- (j);
            \draw [->] (i) -- (k);
            \draw [->] (a) -- (l);

        \end{tikzpicture}
        \caption{}
    \end{subfigure}
    \begin{subfigure}{.42\textwidth}
        \centering
        \begin{tikzpicture}[thick]
            \draw [color=orange,fill=orange!50,very thin]
                            (-0.5,-6*0.8) circle[radius=12pt];
            \draw [color=cyan,fill=cyan!50,very thin]
                            (2.5,-6*0.8) circle[radius=12pt];
            \draw [color=violet,fill=violet!50,very thin]
                            (3.5,-2*0.8) circle[radius=12pt];

            \node (a) at (0, 0*0.8) [vertex] {$a$};
            \node (b) at (-0.5,-2*0.8) [vertex] {$b$};
            \node (c) at (-1,-4*0.8) [vertex] {$c$};
            \node (d) at (-1.5,-6*0.8) [vertex] {$d$};
            \node (e) at (-0.5,-6*0.8) [vertex] {$e$};
            \node (f) at (1,-4*0.8) [vertex] {$f$};
            \node (g) at (0.5,-6*0.8) [vertex] {$g$};
            \node (h) at (2.5,-2*0.8) [vertex] {$h$};
            \node (i) at (2,-4*0.8) [vertex] {$i$};
            \node (j) at (1.5,-6*0.8) [vertex] {$j$};
            \node (k) at (2.5,-6*0.8) [vertex] {$k$};
            \node (l) at (3.5,-2*0.8) [vertex] {$l$};
            \node (p1) at (-1.5,-3*0.8) {$p_1$};
            \node (p2) at (-0.5,-5.25*0.8) {$p_2$};
            \node (p3) at (0.25,-4*0.8) {$p_3$};
            \node (p4) at (2.75,-4*0.8) {$p_4$};
            \node (p5) at (3.25,-6*0.8) {$p_5$};
            \node (p6) at (3.5,-2.75*0.8) {$p_6$};

            \draw [->,line width=1pt]     (a) -- (b);
            \draw [->,line width=1pt]     (b) -- (c);
            \draw [->,line width=1pt]     (c) -- (d);
            \draw [->][dashed]  (c) -- (e);
            \draw [->][dashed]  (b) -- (f);
            \draw [->,line width=1pt]     (f) -- (g);
            \draw [->][dashed]  (a) -- (h);
            \draw [->,line width=1pt]     (h) -- (i);
            \draw [->,line width=1pt]     (i) -- (j);
            \draw [->][dashed]  (i) -- (k);
            \draw [->][dashed]  (a) -- (l);

            \begin{pgfonlayer}{background}
                \fill[blue,opacity=0.50] \convexpath{a,b,c,d,c,b,a}{12pt};
                \fill[green,opacity=0.50] \convexpath{f,g,f}{12pt};
                \fill[yellow,opacity=0.50] \convexpath{h,i,j,i,h}{12pt};

            \end{pgfonlayer}
        \end{tikzpicture}
        \caption{}
    \end{subfigure}
    \begin{subfigure}{.23\textwidth}
        \centering
        \begin{tikzpicture}[thick]
            \node (1) at (0*0.8,0*0.8) [vertex][fill=blue!50] {$p_1$};
            \node (2) at (-1.5*0.8,-2*0.8) [vertex][fill=orange!50] {$p_2$};
            \node (3) at (-0.5*0.8,-2*0.8) [vertex][fill=green!50] {$p_3$};
            \node (4) at (0.5*0.8,-2*0.8) [vertex][fill=yellow!50] {$p_4$};
            \node (5) at (0.5*0.8,-4*0.8) [vertex][fill=cyan!50] {$p_5$};
            \node (6) at (1.5*0.8,-2*0.8) [vertex][fill=violet!50] {$p_6$};
            \node (g) at (0.5*0.8,-6.2*0.8) [emptyVertex] {};

            \draw [->] (1) -- (2);
            \draw [->] (1) -- (3);
            \draw [->] (1) -- (4);
            \draw [->] (4) -- (5);
            \draw [->] (1) -- (6);

        \end{tikzpicture}
        \caption{}
    \end{subfigure}
    \caption{DFS Tree $T$, Heavy Light decomposition with paths in $\mathcal{P}$, and the corresponding Shallow Tree.}
    \label{fig:hld}
    \vspace{-2mm}
\end{figure*}
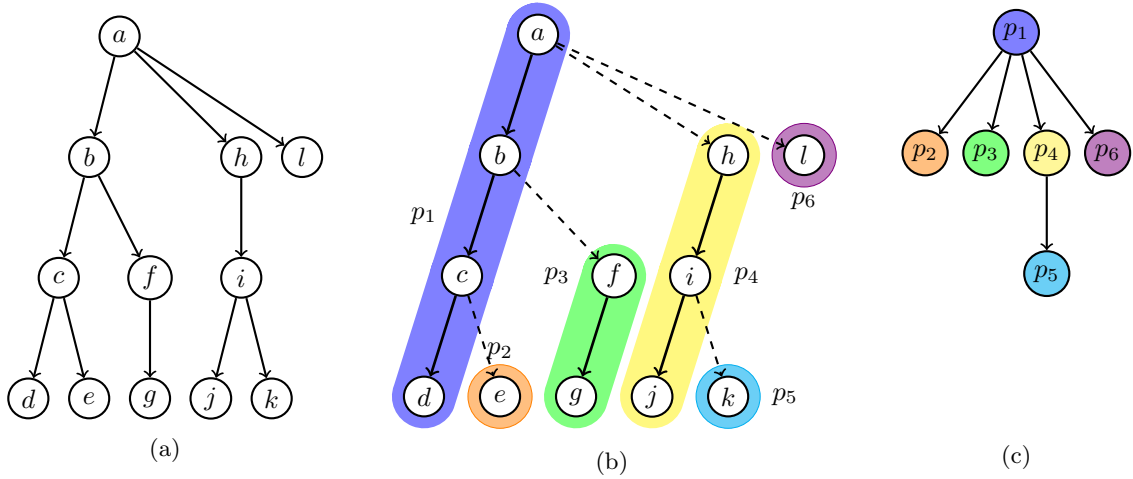

Figure \ref{fig:hld} demonstrates how a DFS tree is decomposed to form a set of ancestor-descendant paths $\mathcal{P}$ which is subsequently used to form the shallow tree $S$. \par

To avoid ambiguity, we address the vertices in the shallow tree as `nodes' and the vertices in the DFS tree as `vertices'. $node\br{x}$ denotes the shallow tree node corresponding to the path in ${\cal P}$ containing vertex $x$.

%
%
The construction of $S$ described above ensures the following simple but crucial properties.
\begin{itemize}
    \item As a result of heavy-light decomposition of a tree $T$ with $n$ vertices, there can be at most $\log{n}$ dashed
        edges on any root to leaf path in $T$. Recall that each edge in $S$ corresponds to a dashed edge. Thus, the depth of any node in tree $S$ can't be larger than $\log{n}$. It is because of this small depth that we choose the name {\em shallow tree} for $S$.
    \item From the DFS property, neighbours of any vertex $v \in V$ are either ancestors or descendants of $v$ in $T$.
        Consider any such neighbour $u$. Let $p_1$ be the path in $\mathcal{P}$ containing $v$ and $p_2$ be the path in ${\cal P}$ containing $u$. $u$ and $v$ may also lie in the same path in $\mathcal{P}$.
        From the construction of $S$, $p_1$ and $p_2$ will share an ancestor-descendant relation in $S$. So we can state the following lemma.
        \begin{lemma}
            \label{lem:anc-des}
            For a DFS tree $T$ of an undirected graph $G = \br{V, E}$ with shallow tree $S$, any vertex $v \in V$ which lies in node $\mu \in S$ can have edges only to vertices lying in the nodes which are ancestors or descendants of $\mu$ in $S$.
        \end{lemma}
\end{itemize}

We require that the vertices of each solid path have consecutive $dfn$.
This enables us to represent each path $p\in \mathcal{P}$ in a compact manner using just the smallest and largest $dfn$ of vertices on $p$. For each path $p$, we store this pair as $PathEndPoints$ at the corresponding node of the shallow tree $S$. Assigning consecutive $dfn$ to each solid path can be accomplished easily - we carry out another DFS on $T$ where for each vertex, the next vertex to be visited is its child hanging through a solid edge (Refer to Algorithm \ref{algo:preprocess-mani} in Appendix).


\section{Rerooting DFS Tree $T$}
\label{sec:reroot}

Given a DFS tree $T$ for a graph $G = \br{V, E}$ and a vertex $r'\in V$, the objective is to compute a DFS tree $T^*$ rooted at $r'$ for the same graph $G$. First we compute the shallow tree representation $S$ of $T$ in $\OO{n}$ time. We now describe the rerooting procedure.

\subsection{Reroot Procedure}
The tree $T^*$ is empty in the beginning and is grown gradually starting from $r'$.
To build this tree efficiently, the idea is to re-use the paths from ${\cal P}$.
In addition, we aim to populate a {\em reduced adjacency list} for each vertex, which should be small and yet sufficient to compute a DFS tree.

The intuition sketched above is materialized by carrying out a DFS traversal {\em guided} by the shallow tree $S$.
Note that a node of the shallow tree corresponds to a path in ${\cal P}$.
To compute $T^*$, our algorithm performs a sequence of steps. Each step begins with entering a node of $S$ through some vertex present on the path stored at the node and leaving it after traversing the path along one direction.
The first node to be visited is $node(r')$. We now provide complete details of the computation involved in each step.
%
%
%
Consider any node $\nu\in S$. Let $path(y,z)$ be the path corresponding to $\nu$.
When the DFS traversal enters $\nu$ through a vertex, say $x$, the following 3 simple operations are carried out.
\begin{enumerate}
    \item \emph{Move towards the farther end of the path.}\\
        We determine the vertex from $\{y,z\}$ farther from $x$. Let this vertex be $y$. DFS traversal proceeds from $x$ to $y$ and $path(x, y)$ is attached to the tree $T^*$.
        Next, we update the $PathEndpoints\br{\nu}$ such that it stores the endpoints of the untraversed part of $path(y,z)$.
    \item \emph{Populate the reduced adjacency list of the path just traversed.}\\
        For vertices on $path(x,y)$, we populate the reduced adjacency list $\mathcal{L}$ using ancestors and descendants of $\nu$ in the shallow tree $S$. This has been explained below in Section \ref{subsubsec:populate}.
    \item \emph{Continue traversal.}\\
        Using the reduced adjacency list $\mathcal{L}$ of the traversed path computed in step 2 above,
        we continue the DFS traversal along the unvisited neighbours of the vertices in the order from $y$ to $x$ (opposite to the direction of traversal, due to the recursive nature of DFS).
\end{enumerate}

Algorithm \ref{algo:reroot} presents the complete pseudocode of the rerooting procedure based on the above 3 steps.
Invoking Reroot$(r')$ produces the DFS tree rerooted at $r'$.
\begin{algorithm*}
    \BlankLine
    \Fn{Reroot (x)} {
        $(y, z) \gets PathEndpoints\br{node\br{x}}$ \;            
        \lIf(\tcc*[f] {compute $dist$ using $dfn$s}){ $dist(x, z) > dist(x, y)$ }{$\text{Swap}\br{y, z}$}
        Attach $path\br{x, y}$ to $T^*$\;
        \If {$ x \neq z $} {
            $ w \gets \text{Neighbour of } x \text{ on }path(y, z)\text{ nearest to }z$ \;
            $ PathEndpoints\br{node\br{x}} \gets \br{w, z}$       \tcc*{untraversed path}
        }

        $\mathcal{L} \gets ReducedAL\br{\mathcal{L}, \br{x, y}}$ \;
        \lFor{$i = \text{dfn}(y)$ \KwTo $\text{dfn}(x)$}{
            $status(v(i)) \gets visited$
        }
        \For{$i = \text{dfn}(y)$ \KwTo $\text{dfn}(x)$}{
            \ForEach{vertex $u \in \mathcal{L}(v(i))$}{
                \lIf{ $status(u) = unvisited$ } {
                    \{add $\br{v\br{i}, u}$ to $T^*$; Reroot($u$)\}
                }
            }
        }
    }
    \caption{Recursive procedure to reroot the DFS Tree $T$}
    \label{algo:reroot}
\end{algorithm*}
\vspace{-2mm}

\subsubsection{Populating Reduced Adjacency Lists}
\label{subsubsec:populate}

Here we define a query $Q\br{u, (p_s, p_e)}$, where $u,p_s,p_e\in V$ satisfy the following constraints:
\begin{itemize}
    \item $p_s$ and $p_e$ are the end-points of an ancestor-descendant path in $T$.
    \item $u$ is a descendant of the highest vertex on the $path\br{p_s, p_e}$, but does not lie on the path.
\end{itemize}
This query returns the edge from vertex $u$ that is incident to the $path(p_s,p_e)$ closest to $p_e$, but returns \textit{null} if no such edge exists.

Consider any node $\nu\in S$, and let $path(y,z)$ be its corresponding path in ${\cal P}$. When the DFS traversal enters $\nu$ through
$x$ and proceeds towards $y$, we populate the reduced adjacency lists of vertices on $path(x,y)$ using Lemma \ref{lem:anc-des} as follows.
\begin{itemize}
    \item \textit{Processing Ancestors.}\\
        For each $u\in path(x,y)$ and for each ancestor $\mu$ of $\nu$, we add $Q\br{u, PathEndpoints\br{\mu}}$ to $\mathcal{L}\br{u}$.
    \item \textit{Processing Descendants.}\\
        Using the shallow tree $S$, one can list all the vertices that are descendants of $path(x,y)$ in $T$. From all these vertices, we query for an edge to $path(x,y)$ which is incident closest to $y$. For any descendant $u$, if the query $Q\br{u, \br{x,y}}$ returns a non-\textit{null} value, say $w$, then we add $u$ to $\mathcal{L}\br{w}$.
\end{itemize}
\begin{remark}
    In Algorithm \ref{algo:reroot}, after visiting $x$, if we moved towards the leaf node, then the untraversed part of $\nu$ has to be treated as an ancestor path while populating the reduced adjacency lists of $path\br{x,y}$, and as a descendant path, otherwise.
\end{remark}
\begin{algorithm*}
    \BlankLine
    \Fn{ReducedAL $\br{\mathcal{L}, \br{x, y}}$}{
        $\mu \gets \text{parent}\br{\text{node}\br{x}}$\;
        \While(\tcc*[f] {edges to ancestor paths}){$ \mu \neq \text{NULL} $}{
            \For{$i = \text{dfn}(y)$ \KwTo $\text{dfn}(x)$}{
                $\mathcal{L}\br{v\br{i}} \gets \mathcal{L}\br{v\br{i}} \cup Q(v\br{i}, PathEndPoints\br{\mu})$\;
            }
            $\mu \gets parent(\mu)$ \;
        }
        \leIf(\tcc*[f]{$z$ is ancestor among two}){$\text{dfn}\br{x} < \text{dfn}\br{y}$}{$z \gets x$}{$z\gets y$}
        $\mathcal{C} \gets \text{Desc}_T\br{z} \setminus \bc{v\br{dfn\br{x}},\dots,v\br{dfn\br{y}}}$\;
        \ForEach(\tcc*[f]{each descendant of $path\br{x, y}$}){$u \in \mathcal{C}$}{
            $w \gets Q\br{u, \br{x, y}}$                                      \tcc*{ensure $w$ is closest to $y$}
            \lIf {$ w \neq \text{NULL} $}{
                $\mathcal{L}\br{w} \gets \mathcal{L}\br{w} \cup \{u\}$
            }
        }
        \Return $\mathcal{L}$
    }
    \caption{Populating reduced adjacency lists of vertices on $path(x,y)$.}
    \label{algo:reducedal}
\end{algorithm*}

In Algorithm \ref{algo:reducedal}, we used query $Q\br{u, \br{p_s, p_e}}$ as a black box.
This query can be answered very easily as follows.
The constraints of $Q\br{u, \br{p_s, p_e}}$ imply that the returned edge is always an edge from $u$ to an ancestor of $u$. For answering this query, we use a data structure $\mathcal{D}$ which stores the following information for each vertex $u$.
\begin{definition}
    \label{def:D}
    $\mathcal{D}\br{u}$ is an array that stores all the ancestors of $u$ in $T$ to which $u$ is a neighbour and it stores them in the increasing order of distance from the root.
\end{definition}

This data structure enables us to answer query $Q\br{u, \br{p_s, p_e}}$ using a binary search on array $\mathcal{D}\br{u}$ and it takes $\OO{\log{n}}$ time only. Interestingly, the data structure $\mathcal{D}$ can be preprocessed very easily in $O(m+n)$ time as follows.
We visit vertices in the increasing order of their depth first numbers. Note that for each vertex $v$, the neighbours of $v$ which have $dfn$ larger than that of $v$ are its descendants. For each such descendant $u$, we append $v$ to $\mathcal{D}(u)$.
Iterating in increasing order of $dfn$ ensures that all arrays in $\mathcal{D}$ are sorted as needed (Refer to Algorithm \ref{algo:preprocess-dfs} in Appendix).
\begin{theorem}
    \label{thm:D}
    Given a DFS Tree $T$ of an undirected graph $G$, there exists a data structure which occupies exactly $m$ words, can be constructed in $\OO{m+n}$ time and can answer the query $Q\br{u, \br{p_s, p_e}}$ in $\OO{\log{n}}$ time.
\end{theorem}

\subsection{Time Complexity Analysis}
\label{ssec:tca}

During preprocessing, we need to build the DFS Tree $T$, carry out heavy-light decomposition to obtain $P$, construct the shallow tree $S$, and build $\mathcal{D}$. This processing as shown earlier can be completed in $\OO{m+n}$ time.
The time complexity of the Reroot procedure is bounded by the time required to populate the reduced adjacency lists $\mathcal{L}$. This in turn, is bounded by the number of calls to query $Q$. \par

To analyse the number of calls made from any vertex $w$, let $\nu$ be the node in the shallow tree $S$, containing $w$.
In general, if the bound on height of $S$ is $d$, ReducedAL makes worst-case $d$ queries from $w$ to ancestors of $\nu$. Also when an ancestor of $\nu$ in $S$, say $\mu$, is visited during Reroot procedure, ReducedAL makes a query from $w$ to $\mu$.
Note that when we enter any path in $\mathcal{P}$, our choice of direction ensures that at least half of the path is traversed. This path halving technique (also used in~\cite{AggarwalAK90,BaswanaCC016}) ensures that any node in $S$ (or a path in {\cal P}) is visited at most $\log{n}$ times.
This implies that any such ancestor $\mu$ may be visited at most $\log{n}$ times. Thus, we can have worst-case $d\br{\log{n} + 1}$ queries from $w$ to its ancestor paths throughout the Reroot procedure.
Summing over all the vertices, there can be at most $nd\br{\log{n}+1}$ calls to query $Q$.
Therefore, populating the reduced adjacency lists $\mathcal{L}$ takes overall $\OO{nd\log^2{n}}$ time.
From section \ref{sec:shallow-tree}, we know $d \leq \log{n}$. Thus, using theorem \ref{thm:D}, we can state the following lemma.
\begin{lemma}
    \label{lem:reroot}
    Given a DFS Tree $T$ of an undirected graph $G$, there exists a data structure of size $\OO{m+n}$ words, which can be constructed in $\OO{m+n}$ time, and it can be used to compute a DFS Tree of $G$ rooted at any given vertex in $\OO{n\log^3{n}}$ time.
\end{lemma}

\subsubsection{Getting rid of a $\log$ factor}

Consider the moment when the Reroot procedure enters a $path\br{y, z}$ through the vertex $x$ and reaches the endpoint $y$.
In Algorithm \ref{algo:reducedal}, for each descendant $w$ of $path\br{x, y}$, we perform query $Q\br{w, \br{x, y}}$ separately.
Instead, using Fractional Cascading, we can perform all these queries together in an efficient manner.
Among $x$ and $y$, let $x$ be the vertex closer to the root of $T$.
As described earlier, all the vertices of $path(x,y)$ have consecutive $dfn$, and so do the vertices of subtree $T\br{x}$.
Let $last\br{x}$ be the vertex in $T\br{x}$ with the largest $dfn$.
Since vertices on $path\br{x, y}$ have been visited, we need to query for edges only from vertices with $dfn$ between $dfn\br{y}+1$ and $dfn\br{last\br{x}}$.
Finding edges to $path\br{x, y}$ from these vertices can be done with a single query to the fractionally cascaded $\mathcal{D}$ (Lemma \ref{lem:FC}).
It takes $\OO{\log{m} + dfn\br{last\br{x}} - dfn\br{y}}$ time to execute this query.
We charge the $\log{m}$ part of the query time to the vertices on $path\br{x, y}$ and the $dfn\br{last\br{x}} - dfn\br{y}$ is distributed among the descendant vertices. Thus, each descendant vertex incurs a constant charge.

Note that, queries to the ancestors of $path\br{x, y}$ are answered using the original $\mathcal{D}$ itself.
Therefore, in a shallow tree of height $d$, each vertex $v \in V$ incurs following charges during Reroot procedure - $\OO{d\log{n}}$ when $v$ acts as a descendant in the queries made while visiting ancestors of $v$, and $\OO{d\log{n} + \log{m}}$ while visiting $v$ itself. Overall the charge on any vertex is $\OO{d\log{n}}$ and, therefore, the time complexity of Reroot procedure reduces to $\OO{nd\log{n}}$.
So we can state the following lemma.
\begin{lemma}
    \label{lem:reroot2}
    Given an undirected graph $G$ and a shallow tree with height $d$, there exists a data structure which occupies $\OO{m+n}$ words, can be constructed in $\OO{m+n}$ time, using which the Reroot algorithm executes in $\OO{nd\log{n}}$ time.
\end{lemma}

%

\vspace{-2mm}
\subsection{Correctness of Reroot Procedure}
As is evident from the pseudocode of Algorithm \ref{algo:reroot}, besides populating the reduced adjacency lists, Reroot can be seen as the usual DFS traversal. So in order to show that Reroot indeed computes a valid DFS tree, all we need to show is that the reduced adjacency list of each vertex is populated {\em correctly}. We can show this as follows. \par
During Reroot procedure, consider the moment when we attach some path, say $path(x,y)$ to $T^*$.
Lemma \ref{lem:anc-des} implies
that all the neighbours of vertices on the $path(x,y)$
will lie either in ancestors or in descendants of $path(x,y)$ in the shallow tree.
For each descendant vertex of $path(x,y)$, we add an edge incident on $path(x,y)$ closest to $y$ (lines 12,13 of Algorithm \ref{algo:reducedal}). For each ancestor path of $path(x,y)$, we
add an edge from each vertex of $path(x,y)$ (line 5 of Algorithm \ref{algo:reducedal}). This ensures that from each connected component in the graph induced by unvisited vertices, the edge incident on $path(x,y)$ closest to vertex $y$ is surely added to $\mathcal{L}$ (if exists).
Hence we can conclude that Reroot procedure indeed computes a DFS tree.

%


\section{Extension to Fault Tolerant DFS Tree}
\label{sec:k-fault}

Let $F$ be the set of failures (edges/vertices) in any undirected graph $G$, with $|F| = k$. Here, we describe how with some elementary modifications, procedure Reroot can be utilized to report DFS tree of $G \setminus F$.
First, we update the set $\mathcal{P}$ and the shallow tree $S$ as follows.
\begin{enumerate}
    \item Each vertex maintains a state: {\em active} or {\em failed}. For a failed vertex $x \in V$, we toggle $x$'s state to \textit{failed}. Let $p \in {\cal P}$ be the path containing $x$. We remove $x$ from $p$. The resulting smaller paths are added to $\mathcal{P}$ and $p$ is removed from $\mathcal{P}$.
    \item For each failed edge $e = (u, v)$, we mark the corresponding entries in the adjacency lists of $u$ and $v$ as {\em failed}. Here, we do not make any changes to fractionally cascaded $\mathcal{D}$. If failed edge $e$ is a tree-edge and was marked \textit{solid} during heavy-light decomposition, $\mathcal{P}$ is updated as follows. Path $p \in \mathcal{P}$ containing $e$ is split into two smaller paths. These smaller paths are added to $\mathcal{P}$ and $p$ is removed from $\mathcal{P}$.
    \item After updating $\mathcal{P}$, shallow tree $S$ is updated as per Definition \ref{def:shallow-tree}. For any path $p \in {\cal P}$, let the vertex in $p$ closest to root of $T$ be $x$. The node in $S$ corresponding to $p$ is attached to $S$ as a child of the node containing the closest active ancestor of $x$.
\end{enumerate}
Given the set $F$, we can update $\mathcal{P}$ and $S$ in $\OO{n}$ time as follows - Make a DFS traversal through $T$ while ensuring vertices are visited in increasing order of their $dfn$. Update the paths and the edges in $S$ as discussed above.
For each failure, the number of nodes in $S$ increases by at most one. After $k$ updates, $S$ can have $k$ new nodes and may not be shallow anymore. Note that the depth of a node $\nu$ in $S$ doesn't increase due to any failure which doesn't lie on the path from $\nu$ to root of $S$. Thus, if the maximum of number of failures on all root-leaf paths in $S$ is $k'$, the height of the shallow tree is at most $k' + \log{n}$.

To ensure that no \textit{deleted} vertices/edges are traversed during Reroot procedure, we make the following modifications.
The result of any query $Q\br{v, \br{p_s, p_e}}$ may be a \textit{failed} edge or an edge to a \textit{failed} vertex. In such a case, we iterate in $\mathcal{D}\br{v}$ towards $p_s$ from the invalid edge until we find an edge present in $G \setminus F$. However, if we cross $p_s$ in doing so, we stop and return $null$. In the worst-case, we spend $\OO{nk'+2k'}$ time in such iterations.
Calling the procedure Reroot after above updates suffices to report the DFS tree of $G \setminus F$.
Using Lemma \ref{lem:reroot2}, we can conclude the following theorem.
\begin{theorem}
    \label{thm:k-fault}
    An undirected graph $G$ and its DFS tree $T$ can be preprocessed in $\OO{m+n}$ time to build a data structure of $\OO{m+n}$ size, using which one can compute the DFS tree of the graph for any given $k$ failed vertices or edges, in $\OO{n\br{k'+\log{n}}\log{n}}$ time, where $k' \leq k$ is the maximum number of faults on any root-leaf path in the tree $T$.
\end{theorem}

\section{Fully Dynamic DFS}
\label{sec:fully-dynamic}

We first describe how the fault tolerant DFS algorithm can handle incremental updates.
Following that, we use the {\em overlapped periodic rebuilding technique} to arrive at a fully dynamic DFS algorithm.
The ideas utilized in both of these steps were used by Baswana \etal.~\cite{BaswanaCC016}.

Let $U$ be the set of updates in any undirected graph $G$.
In order to handle each vertex insertion, we add the new vertices in $V$ and add the corresponding edges to $E$ and $\mathcal{L}$.
For the edges insertions, we directly add these edges to reduced adjacency list of endpoints of the edge. These modifications are sufficient to handle incremental updates.
Since the size of reduced adjacency lists after these updates is $\OO{n\br{|U|+\log{n}}\log{n}}$, we get the same worst-case time bound for the time complexity of Reroot procedure.

The following lemma helps us formulate the fully dynamic DFS algorithm from the fault tolerant one.

\begin{lemma}
    (\textit{Lemma 6.1 in \cite{BaswanaCC016}}) Let $\mathcal{D}$ be a data structure that can be used to report the solution of a graph problem after a set of $U$ updates on an input graph $G$. If $\mathcal{D}$ can be built in $O(f)$ time and the solution for graph $G+U$ can be reported in $O(h+|U|\cdot g)$time, then $\mathcal{D}$ can be used to report the solution after every update in worst-case $O(\sqrt{fg}+h)$ update time, given that $f/g \le n$.
\end{lemma}
Substituting $f = m$, $g = n \log{n}$ and $h = n \log^2{n}$, we obtain, $\OO{\sqrt{fg}+h}$ = $\OO{\sqrt{mn\log{n}}}$. Hence this implies the following theorem:

\begin{theorem}
    \label{thm:dynamic}
    An undirected graph can be preprocessed in $\OO{m+n}$ time to build a data structure occupying $\OO{m+n}$ words, using which one can maintain a DFS tree for any online sequence of insertions and deletions of vertices/edges in $\OO{\sqrt{mn\log{n}}}$ worst-case time per update.
\end{theorem}

\section{Conclusion}

We presented a drastically simpler algorithm for DFS in an undirected graph in fault tolerant setting. This algorithm takes $O(n(k'+\log n)\log n)$ time for reporting a DFS tree for any given set of $k$ failed vertices/edges, where $k'$ is the maximum number of failed vertices/edges along any root-leaf path of the initial DFS tree. This is superior to all the previous bounds on this problem. Moreover,
we achieve optimal space and optimal preprocessing time. 

Like each of the previous algorithms on this problem, our algorithm is extendible to the case when we have insertion of vertices as well. For the generic setting where $k$ updates may be insertion or deletion of vertices, note that the input size of the query itself can take $\Theta(nk)$ space - insertion of $k$ vertices each with $\Theta(n)$ edges. 
So our bounds may just be nearly optimal for this generic setting. However, there may exist a faster algorithm
for the restricted case, especially when the updates are only insertion or deletion of edges. 
It will be interesting to design faster decremental and/or fully dynamic algorithm for DFS (when the updates are only edges). We  feel that the simplicity of the algorithm presented in this paper will surely pave way for designing these algorithms. 

\bibliography{src/references}

\begin{appendices}
    



\section{Applications of Fully Dynamic DFS}
\label{appl}

Baswana et al.~\cite{BaswanaCC016} in the full version of their paper show how their algorithm can be used to solve various dynamic subgraph problems such as dynamic subgraph connectivity, biconnectivity, and 2-edge connectivity. In this section, we state these problems and show how the new algorithm can be used to improve the worst case time complexity.

\subsection{Dynamic Subgraph Connectivity}
\subsubsection*{Problem definition \& Existing results}
Given an undirected graph, the goal is to process an input stream of updates and queries as efficiently as possible. Each update toggles the state of some vertex in the graph from {\em active} to {\em inactive} or vice-versa. The queries check if any given pair of vertices are connected in the subgraph induced by active vertices.
Chan~\cite{Chan06} formally introduced this problem in 2006. They presented an algorithm with $\OO{m^{0.94}}$ amortized update time that could answer connectivity queries in $\tilde{\cal O}\br{m^{1/3}}$ time. Later in 2008, Chan \etal.~\cite{ChanPR08} presented a new algorithm that improves the update time to $\tilde{\cal O}\br{m^{2/3}}$.

Holm \etal.~\cite{HolmLT01} in 1998 presented a deterministic algorithm for fully dynamic connectivity on edge updates which takes amortized $\tilde{\cal O}\br{1}$ time per update and $\tilde{\cal O}\br{1}$ time per query. For the same problem, Kapron \etal.~\cite{KapronKM13} presented a Monte Carlo algorithm with poly-logarithmic worst case bounds in 2013.
Since switching the state of a vertex is equivalent to $\OO{n}$ edge updates, the algorithms by Holm \etal~\cite{HolmLT01} and Kapron \etal~\cite{KapronKM13} can be used for the dynamic subgraph connectivity problem with the respective update times multiplied by a factor of $n$.

The DFS-tree based algorithm by Baswana \etal.~\cite{BaswanaCC016} was the first deterministic algorithm for dynamic subgraph connectivity with $o\br{m}$ worst case update time and a constant query time. In addition to switching the state of existing vertices, it allows insertion of new vertices as well. Their algorithm processes each update in $\OO{\sqrt{mn}\log^{2.5}{n}}$ and answers each query in $\OO{1}$ time.
The subsequent results by Chen \etal.~\cite{ChenDWZ16} and Nakamura and Sadakane~\cite{NakamuraS17} for updating DFS trees also improve the worst case update time and space occupied by the data structure for the given problem.
Using our dynamic DFS Tree algorithm, we extend the results of~\cite{BaswanaCC016,ChenDWZ16,NakamuraS17} by improving the worst-case time complexity to $\OO{\sqrt{mn\log{n}}}$.

\subsubsection*{Our Algorithm}

We maintain the DFS tree of the graph $G$. This tree is rooted at a dummy vertex $r$ and other trees in the forest are hanging from the vertex $r$. To check if any two vertices are connected in $G$, we find the lowest common ancestor of the two vertices in the DFS tree. If the required ancestor is $r$, then the pair is not connected in $G$. If any other vertex is the lowest common ancestor, then the pair is connected. After each update, the DFS tree can be refreshed in $\OO{\sqrt{mn\log{n}}}$ time. This also enables us to answer the queries in $\OO{1}$ time.

\subsection{Dynamic Subgraph Biconnectivity and Subgraph 2-edge Connectivity}
\subsubsection*{Problem definition \& Existing results}
The structure and information stored in DFS trees also helps in solving dynamic subgraph biconnectivity and subgraph 2-edge connectivity problems.
Baswana \etal.~\cite{BaswanaCC016} presented an $\OO{\sqrt{mn}\log^{2.5}n}$ worst case update time and constant query time algorithm by augmenting their dynamic DFS algorithm.
We demonstrate how the new algorithm can be augmented on similar lines to solve these problems as well.

A {\em Biconnected Component} in a graph $G$ is a maximal subgraph $S$ with the following property - On failure of any vertex $v$ within $S$, subgraph $S \setminus \{v\}$ remains connected. Similarly, a {\em 2-edge Connected Component} is a maximal subgraph $S$ with the property - after failure of any edge $e$ in the graph $G$, vertices in $S$ remain connected.
In both the problems, the input graph is partitioned into subgraphs, where each subgraph corresponds to one such component. A query involves checking if a given pair of vertices lie within the same component.

Another perspective to this partitioning is through {\em Articulation points} and {\em Bridges}.
\begin{definition}
    \label{def:articulation}
    Given an undirected graph $G = (V, E)$, a vertex $x \in V$ is an articulation point iff there exists a pair of vertices $u$ and $v$, such that every path from $u$ to $v$ passes through $x$.
\end{definition}

\begin{definition}
    \label{def:bridge}
    Given an undirected graph $G = (V, E)$, an edge $e \in E$ with endpoints $x, y \in V$ is a bridge iff there doesn't exist any path between $x$ and $y$ but the one through edge $e$.
\end{definition}

Articulation Points and Bridges act as boundaries between biconnected or 2-edge components, and thus partition the graph into a set of connected components. Articulation Points and Bridges can be found using {\em Depth First Numbers}.
Using these DFNs, the {\em high-point} of each vertex is defined. The {\em high-point} of any vertex $x$ is the smallest DFN $a$, such that either $dfn\br{x} = a$ or there is a back edge from $x$ or descendant of $x$ incident on a vertex $w$ such that $dfn\br{w} = a$. Any non-root vertex $x \in V$ is an articulation point if the high-point of any child of $x$ is $ \ge dfn\br{x}$. The root vertex is an articulation point if has two or more children in $T$. Any tree edge $(x, y)$, with $x = par_T(y)$, is a bridge iff the high-point of $y$ is $dfn\br{y}$. Thus, one can find the biconnected or 2-edge connected components in $\OO{m+n}$ time using DFS traversal. Moreover, if one has the DFS tree $T$ and high-points of all the vertices, then the components can be computed in $\OO{n}$ time. \par

\subsubsection*{Our Algorithm}
We augment the new dynamic DFS algorithm such that it can compute the high-points of all vertices while computing the new DFS tree $T^*$. To compute the high-points of any vertex $x$, the static algorithm scans over the edges incident on $x$. Instead of scanning over all the edges, we build a compact list of edges that are sufficient to find the high-points. This list, let's call it $A$, is built while constructing the DFS Tree $T^*$ and satisfies the following crucial property:


\textit{ For every edge $(x, y)$ in the updated graph, where $x$ is an ancestor of $y$ in $T^*$, at-least one edge $(u,v)$ is present in $A$ such that $u$ is $x$ or an ancestor of $x$ in $T^*$ and $v$ is $y$ or a descendant of $y$ in $T^*$.}

List $A$ can be populated by incorporating the following to the Reroot procedure.
First, all new edges added to graph $G$ are added to $A$. Now during the Reroot procedure, when the traversal enters the shallow tree node $\nu$ at vertex $v$, and travels towards the farther endpoint, say vertex $y$, following edges are added to $A$. Let $t$ be the vertex closest to root of $T$ among $v$ and $y$.
\begin{enumerate}
    \item For every unvisited descendant $d$ of $t$, the edge from $d$ to path $v \rightarrow y$ incident closest to $v$, if any, is added.
    \item For every visited descendant $d$ of $t$, the edge from $d$ to path $v \rightarrow y$ incident closest to $y$, if any, is added.
\end{enumerate}
Let's look at the time required to do the above. We need to iterate over all descendants of vertex $t$ and find appropriate edges from descendants to path$\br{v,y}$ as according to above. This is very similar to handling descendants while populating the reduced adjacency lists. Thus it takes 2 queries to ${\cal D}$ to add the required edges to $A$. This doesn't change the time complexity of Reroot procedure. Moreover, instead of creating another list $A$, we could simply add these edges to the reduced adjacency lists ${\cal L}$ without increasing the size of ${\cal L}$ by anything more than a constant.

Using these lists, we can find the high-points of all vertices in $\OO{n}$ time and thus update the components. The time required per update is $\OO{\sqrt{mn\log{n}}}$, which improves by a $log$ factor on the previous best by Chen \etal.~\cite{ChenDWZ16}. Also, compared to the augmentation used by~\cite{BaswanaCC016}, the changes which we introduce are simple and minimal.

\begin{theorem}
    \label{thm:applications}
    An undirected graph can be preprocessed in $\OO{m+n}$ time to build a data structure of $\OO{m+n}$ words, which can be used to update the solutions for the dynamic subgraph connectivity problem, dynamic subgraph biconnectivity problem, and subgraph 2-edge connectivity problem in $\OO{\sqrt{mn\log{n}}}$ time and to answer each query in $\OO{1}$ time.

\end{theorem}

\begin{remark}
    \label{remark:applications}
    There have been a couple of new results for dynamic subgraph connectivity problems after~\cite{BaswanaCC016} which improved the time bound in some special cases.
    Nakamura~\cite{Nakamura17} presented algorithms for dynamic subgraph connectivity which achieve better update time when the number of leaf nodes in DFS Tree is $o\br{n}$.
    They presented an algorithm that achieves the amortized update time $\OO{\sqrt{m}\log^{1.25}{n} + l\frac{\log^2{n}}{\log{\log{n}}} + n}$ and they also presented a Monte Carlo algorithm with worst-case $\OO{\sqrt{ml}\log^{2.75}{n} + n}$ update time. Here, $l$ is the number of leaf nodes in the DFS tree of the input graph.

    Another notable result is by Nanongkai \etal.~\cite{NanongkaiSW17} in 2017. They presented a Las Vegas algorithm to maintain minimum spanning forest which takes $\OO{n^{o\br{1}}}$ worst-case time(w.h.p.) to process general edge updates. This implicitly results in a Las Vegas algorithm for dynamic subgraph connectivity with $\OO{n^{1+o\br{1}}}$ time per vertex update.
\end{remark}

\section{Preprocessing to make $S$ and $D$}

\begin{algorithm*} 
    \BlankLine
    \Fn{DFS (node)} {
        $status(node) \gets $ visited\;
        $size(node) \gets 1$\;
        \For{$ v \in \mathcal{N}\br{node} $} {
            \lIf{ status$\br{v}$ is unvisited } {$D\br{v} \gets D\br{v} \cup \{node\}$}
        }
        \For{$ v \in \mathcal{N}\br{node} $} {
            \If{ $status(v) = unvisited$ } {
                $par_T\br{v} \gets node$\;
                $DFS(v)$\;
                $size(node) \gets size(node) + size(v)$\;
            }
        }
        $heavy\_child(node) \gets \text{child with heaviest subtree}$             \tcc*{Null if leaf-node}
    }
    \caption{To generate a DFS tree, get sizes of subtrees and find the heaviest child}
    \label{algo:preprocess-dfs}
\end{algorithm*}

\begin{algorithm*}
    \BlankLine
    \Fn{Manipulate(v)} {
        $status(v) \gets visited$\;
        $dfn(v) \gets curr\_time++$\;
        \eIf{$v$ is non-leaf} {
            stnode$\br{v} \gets \text{Manipulate}\br{heavy\_child\br{v}}$\;
        }{
            stnode$\br{v} \gets \text{path}\br{v, v}$\;
            $\mathcal{P} \gets \mathcal{P} \cup \{\text{stnode}\br{v}\}$            \tcc*{New path with single vertex}
        }
        Update $PathEndPoints$ for stnode$\br{v}$\;
        \ForEach(*\tcc*[f]{Iterate in order similar to DFS}){$ x \in \mathcal{N}(v)$} {
            \If{$status(x)$ is unvisited} {
                $ par_S\br{\text{Manipulate}\br{x}} \gets \text{stnode}(v)$         \tcc*{Path with dashed edge}
            }
        }
        \Return $\text{stnode}(v)$
    }
    \caption{Manipulates the $dfn$ and finds the endpoints of paths in $\mathcal{P}$}
    \label{algo:preprocess-mani}
\end{algorithm*}

\end{appendices}

\end{document}